\documentstyle[12pt,epsfig]{article}
 \newcommand{\be}{\begin{eqnarray}}
 \newcommand{\ee}{\end{eqnarray}}
 \newcommand{\beq}{\begin{equation}}
 \newcommand{\eeq}{\end{equation}}
 \newcommand{\ba}{\begin{array}{1}}
 \newcommand{\ea}{\end{array}}
 \newcommand{\bb}{\begin{thebibliography}}
 \newcommand{\eb}{\end{thebibliography}}

 \newcommand{\abstitle}[1]{{\small {\bf #1}}}
 \newcommand{\absauthor}[1]{{\small {\bf #1}}}
 \newcommand{\address}[1]{{\it #1}}
 
\begin{document}
 \begin{center}
 \abstitle{Role of gluons in soft and semi-hard multiple hadron production
in $pp$ collisions at LHC}\\
 \vspace{0.3cm}
 \absauthor{V.A.Bednyakov$^1$, A.A.Grinyuk$^1$, G.I.Lykasov$^{1,4},$  
M.Poghosyan$^{2,3}$}\\ [2.0mm]
 \address{$^1$ JINR, Dubna, Moscow region, 141980,  Russia, \\ 
$^2$ Universit\'a  di Torino/INFN, 10125 Torino, Italy \\
$^3$ CERN, Geneva, Switzerland \\
$^4$ E-mail:~lykasov@jinr.ru
}
 \end{center}
 \vspace{0.1cm}
%Version of April 27, 2009\\
 \vspace{0.2cm} 
{\bf Abstract}
 \vspace{0.1cm}

Hadron inclusive spectra  in $pp$ collisions 
are analyzed within the modified quark-gluon string model including
both the longitudinal and transverse motion of quarks in the proton 
in the wide region of initial energies. The self-consistent analysis
shows that the experimental data on the inclusive spectra of light 
hadrons like pions and kaons at ISR energies can be satisfactorily described 
at transverse momenta not larger than 1-2 GeV/c. We discuss some difficulties 
to apply this model at energies above the ISR and suggest to include the 
distribution of gluons in the proton unintegrated over the internal transverse 
momentum. It leads to an increase in the inclusive spectra of hadrons and
allows us to extend the satisfactory description of the data in the central rapidity 
region at energies higher than ISR. 

\section{Introduction}
 A rather successful description of various characteristics of hadroproduction
processes at not large transfer can be obtained by using the approach for describing 
the soft hadron-nucleon, hadron-nucleus and nucleus-nucleus interactions
at high energies  based on the topological 
$1/N$ expansion in QCD \cite{tHooft:1974,Veneziano:1975}, where $N$ is the number of 
flavours or colours, for example, the quark-gluon string model (QGSM) \cite{kaid1,kaid2},
the VENUS model \cite{Werner:1993}, the dual parton model (DPM) \cite{capell1,capell2},
the coloured-tube models \cite{Casher1,Gurvich} and others.
%\cite{tHooft:1974,Veneziano:1974,Veneziano:1975,Chew:1978}, 
The conventional QGSM and DPM models used the parton distribution functions
(PDF) and the fragmentation functions (FF) integrated over the internal transverse
momenta of partons. The modification of the QGSM including the transverse motion of quarks 
in the initial hadron has been developed in \cite{Veselov:1985} and \cite{LS:1992,LAS}.
It allowed us to describe the inclusive spectra of hadrons produced in $pp$ collisions
as a function of the Feynman variable $x$ and the hadron transverse momentum $p_t$. However,
up to now there has not been a self-consistent analysis of these spectra within the QGSM in the wide
region of initial energies from the ISR to the LHC ones. In this paper we present the results
of the detailed analysis of the inclusive spectra of light hadrons like pions and kaons produced
in $pp$ collisions at ISR energies within the modified QGSM including the internal transverse 
motion of partons in the initial proton. Then we analyze similar spectra of charged hadrons
produced in central $pp$ collisions at initial energies from 500 GeV up to 7 TeV and compare
them with the S$p{\bar p}$S, Tevatron and latest LHC data. We discuss some difficulties to apply the
modified QGSM to the description of the latest data. To avoid these difficulties we suggest to include
the so-called unintegrated gluon distributions in the proton by analyzing soft hadron production
in $pp$ collisions at very high energies. 
 
As is well known, hard processes involving incoming protons, such as deep-inelastic lepton-proton
scattering (DIS), are described using the scale-dependent PDFs.
A distribution like this is usually calculated as a function of the longitudinal momentum fraction $x$
and the square of the four-momentum transfer $q^2=-Q^2$, integrated over the parton transverse momentum $k_t$.
However, for semi-inclusive processes, such as inclusive jet production in DIS,
electroweak boson production \cite{Ryskin:2003}, etc., the parton distributions unintegrated over  
the transverse momentum $k_t$ 
are more appropriate. The theoretical analysis of the unintegrated quark and gluon PDFs can be found, for
example, in \cite{Nikolaev:2002,Ryskin:2010}. According to  \cite{Ryskin:2010}, the gluon distribution function 
$g(k_t)$ at fixed $Q^2$ has the very interesting behaviour at small $x\leq 0.01$, it increases
very fast starting from almost zero values of $k_t$. In other words, $g(k_t)$ in some sense blows up when
$k_t$ increases and then it decreases at $k_t$ close to 100 GeV$/c$.  In contrast, the quark distribution $q(k_t)$ 
is almost constant 
in the whole region of $k_t$ up to $k_t\sim$ 100 GeV$/c$ and much smaller
than $g(x)$.
This parametrization of the PDFs was obtained in \cite{Ryskin:2010} within the leading order (LO) and next-to-leading 
order of QCD (NLO) 
at $Q^2=10^2$ (GeV$/c$)$^2$ and $Q^2=10^4$(GeV$/c$)$^2$ from the known
(DGLAP-evolved \cite{{DGLAP}}) parton densities determined from the global data analysis. 
At small values of $Q^2$ the nonperturbative effects should be included to get the PDFs.
The nonperturbative effects can arise from the complex structure of the QCD vacuum, see for example
\cite{ShVZ:1979}-\cite{Dorokh_Mikh}. For example, the instantons are some of the
well-studied topological fluctuations of the vacuum gluon fields, see for example
\cite{Shuryak:1998}-\cite{Kochelev:1998}
and references therein. In particular, it is shown \cite{Kochelev:1998} that the inclusion of the instantons
results in the anomalous chromomagnetic quark-gluon interaction (ACQGI) which, for the massive quarks, gives 
the spin-flip part of it. Within this approach the very fast increase of the unintegrated 
gluon distribution function at $0\le k_t\le 0.5$ GeV$/c$ and $Q^2=1$ GeV$/c$ is also shown.
These results stimulated us to assume, that the unintegrated gluon distribution in the proton can be included by
analyzing also the soft hadron production in $pp$ collisions. We discuss this possibility at the end
of this paper. 
\section{Inclusive spectra of hadrons in $pp$ collisions}
\subsection{QGSM}
Let us analyze the hadron production in $pp$ collisions
within the QGSM \cite{kaid1,kaid2} or the dual parton model (DPM) \cite{capell1,capell2} including the transverse 
motion of quarks and diquarks in colliding protons \cite{LS:1992,LAS}. As is known, the cylinder-type graphs
presented in Fig.1 make the main contribution to this process. The physical meaning of the graph presented in
Fig.1 is the following. The left diagram of Fig.1, the so-called one-cylinder graph, corresponds to the
case when two colorless strings are formed between the quark/diquark ($q/qq$) and the diquark/quark ($qq/q$) 
in colliding protons, then, after their break, $q{\bar q}$ pairs are created and fragmented to a hadron. 
The right diagram of Fig.1, the so called multi-cylinder graph, corresponds to a creation of the same two 
colourless strings and many strings between sea quarks/antiquarks $q/{\bar q}$ and sea antiquarks/quarks
${\bar q}/q$ in the colliding protons.
   \begin{figure}[h!]
 %\rotatebox{270}
  {\epsfig{file=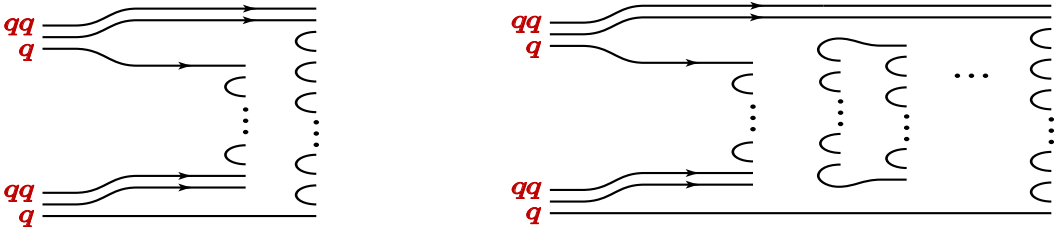,height=3.cm,width=14.cm  }}
  \caption[Fig.2]{The one-cylinder graph (left) and the multicylinder graph (right)
  for the inclusive $p p\rightarrow h X$ process.} 
  \end{figure}
The general form for the invariant inclusive hadron spectrum within the QGSM is the following \cite{kaid1,kaid2,Kaid1}:
\be
E\frac{d\sigma}{d^3{\bf p}}\equiv \frac{2E^*}{\pi\sqrt{s}}\frac{d\sigma}{dxdp_t^2}=
\sum_{n=1}^\infty \sigma_n(s)\phi_n(x,p_t)~,
\label{def:invsp}
\ee
where $E,{\bf p}$ are the energy and three-momentum of the produced hadron $h$ in the l.s. of colliding
protons respectively; $E^*,s$ are the energy of $h$ and the square of the initial energy in the c.m.s of
$pp$; $x,p_t$ are the Feynman variable and the transverse momentum of $h$; $\sigma_n$ is the cross
section for production of the $n$-pomeron chain (or $2n$ quark-antiquark strings) decaying into hadrons, 
calculated within the ``eikonal approximation'' \cite{Ter-Mart}, see Appendix; the function
$\phi_n(x,p_t)$ has he following form \cite{LS:1992} and \cite{MS:2000}-\cite{BLL:2010}:
\be
\phi_n(x,p_t)=\int_{x^+}^1dx_1\int_{x_-}^1 dx_2\psi_n(x,p_t;x_1,x_2)~,
\label{def:phin}
\ee
where
\be
\psi_n(x,p_t;x_1,x_2)=F_{qq}^{(n)}(x_+,p_t;x_1)F_{q_v}^{(n)}(x_-,p_t;x_2)/F_{q_v}^{(n)})(0,p_t)+~\\
\nonumber
+ F_{q_v}^{(n)}(x_+,p_t;x_1)F_{qq}^{(n)}(x_-,p_t;x_2)/F_{qq}^{(n)})(0,p_t)+~\\
\nonumber
2(n-1)F_{q_s}^{(n)}(x_+,p_t;x_1)F_{{\bar q}_s}^{(n)}(x_-,p_t;x_2)/F_{q_s}^{(n)})(0,p_t)~.
\label{def:psin}
\ee
and $x_{\pm}=0.5(\sqrt{x^2+x_t^2}\pm x), x_t=2\sqrt{(m_h^2+p_t^2)/s}$,
\be
F_\tau^{(n)}(x_\pm,p_t;x_{1,2})=\int d^2k_t{\tilde f}_\tau^{(n)}(x_\pm,k_t){\tilde G}_{\tau\rightarrow h}
\left(\frac{x_\pm}{x_{1,2}},{\tilde k}_t;p_t)\right)~,
\label{def:Ftaux}
\ee
\be
F_\tau^{(n)}(0,p_t)=\int_0^1 dx^\prime d^2k_t {\tilde f}_\tau^{(n)}(x^\prime,k_t)
{\tilde G}_{\tau\rightarrow h}(0,p_t)={\tilde G}_{\tau\rightarrow h}(0,p_t)~.
\label{def:Ftauzero}
\ee
Here $\tau$ means the flavour of the valence (or sea) quark or diquark, ${\tilde f}_\tau^{(n)}(x^\prime,k_t)$
is the quark distribution function depending on the longitudinal momentum fraction $x^\prime$ and the 
transverse momentum $k_t$ in the $n$-pomeron chain; ${\tilde G}_{\tau\rightarrow h}(z,{\tilde k}_t;p_t)=
z{\tilde D}_{\tau\rightarrow h}(z,{\tilde k}_t;p_t)$, ${\tilde D}_{\tau\rightarrow h}(z,{\tilde k}_t;p_t)$ is 
the fragmentation function of a quark (antiquark) or diquark of flavour $\tau$ into a hadron $h$.  
We present the quark distribution in a proton in the factorized form ${\tilde f}_\tau(x,k_t)=f_\tau(x)g_\tau(k_t)$,
and choose the $k_t$ distribution of quarks and the FF in the simple Gaussian form
$g(k_t)=(\gamma_q/\pi)exp(-\gamma_q k_t^2)$,
$g_{q\rightarrow h}({\tilde k}_t)=(\gamma_F/\pi)exp(-\gamma_F {\tilde k}_t^2)$, where 
${\tilde k}_t=p_t-zk_t,~ z=x_\pm/x_{1,2}$.
Then, the quark functions in the $n$-pomeron chain will be factorized too \cite{LS:1992} 
\be
{\tilde f}_\tau^{(n)}(x,k_t)=f_\tau^{(n)}(x)g_\tau^{(n)}(k_t)~,
\label{def:ftaun}
\ee
where
\be
g_\tau^{(n)}(k_t)~=~\frac{\gamma_n}{\pi}\exp(-\gamma_n k_t^2), ~ \gamma_n=\frac{\gamma_q}{n}~.
\label{def:gtaun}
\ee
The fragmentation function also reads
\be
{\tilde G}_{\tau\rightarrow h}(z,{\tilde k}_t;p_t)=G_{\tau\rightarrow h}(z,p_t)
g_{\tau\rightarrow h}({\tilde k}_t)~.
\label{def:tGtauh}
\ee
Then, substituting Eqs.(\ref{def:ftaun}) into Eq.(\ref{def:Ftaux}) we get the following form
for $F_\tau^{(n)}$:
\be
F_\tau^{(n)}(x_\pm,p_t;x_{1,2})=f_\tau^{(n)}(x_{1,2})G_{\tau\rightarrow h}(z)
{\tilde I}_n(z,p_t)~,
\label{def:finFtaun}
\ee
where the function ${\tilde I}_n(z,p_t)$ reads \cite{LAS,LS:1992}
\be
{\tilde I}_n(z,p_t)=\frac{\gamma_z}{\pi}\exp(-\gamma_z p_t^2),~
\gamma_z=\frac{\gamma_F}{1+n\rho z^2},~ \rho=\frac{\gamma_F}{\gamma_q}~.
\label{def:gammaz}
\ee
When we take the $k_t$ distribution of quarks (diquarks) $g(k_t)$ and the FF $g_{q\rightarrow h}({\tilde k}_t)$
in the exponential form
\be
g(k_t)=\frac{B_q^2}{2\pi}\exp(-B_q k_t)~,~
g_{q\rightarrow h}({\tilde k}_t)=\frac{B_F^2}{2\pi}\exp(-B_F {\tilde k}_t)~,
\label{def:gkt}
\ee
we get the following form for the function $I_n(z,p_t)$ entering into
$F_\tau^{(n)}(x_\pm,p_t;x_{1,2})$
\be
F_\tau^{(n)}(x_\pm,p_t;x_{1,2})=f_\tau^{(n)}(x_{1,2})G_{\tau\rightarrow h}(z)
I_n(z,p_t)~,
\label{def:finFtaunexp}
\ee
\be
I_n(z,p_t)=C_n
%\left\(\frac{B_q^2}{2\pi}\right\)^n\frac{B_c^2}{2\pi}
\int\frac{J_0(b p_t)bdb}{(z^2b^2+B_q^2)^{3n/2}(b^2+B_F^2)^{3/2}}~.
%I_n(z,p_t)=\frac{B_z^2}{2\pi(1+B_zm_h)}\exp\left(-B_z(m_{ht}-m_h)\right)~.
\label{def:Fnexp}
\ee
Here $J_0(b p_t)$ is the zero order Bessel function and the coefficient $C_n$ is determined by the 
normalization equation
\be
\int d^2 p_t I_n(z,p_t)=1
\label{def:normIn}
\ee 
%Here 
%\be
%B_z=\frac{B_0}{1+n\rho z^2}~.
%          \ee 
Equation (\ref{def:normIn}) is similar to the normalization equation for the function ${\tilde I}_n(z,p_t)$
given by Eq.(\ref{def:gammaz}), when the quark distributions and the FF are chosen as the Gaussian forms.

%%%%%%%%%%%%%%%%%%%%%%%%%%%%%%%%%%%%%%%%%%%%%%%%%%%%%%%%%%%%%%%%%%%%%%%%%%%%%%%%%%%%%%%%%%%%%%%%%%%%%%%%%%
$\bullet~${\bf Gluon distribution in proton }.\\
As is mentioned in the Introduction, the unintegrated gluon distribution in the proton at small values of
$x$, as a function of $k_t$, increases very fast when $k_t$ increases and then slowly decreases, according
to the instanton vacuum approach for the massive quarks \cite{Kochelev:1998} at small $Q^2\sim$ 1 GeV$/c$.
The similar behaviour for $g(k_t)$ was obtained within the NLO QCD calculations at large $Q^2=10^2$GeV$/c$
and $Q^2=10^4$GeV$/c$ for the massless quarks \cite{Ryskin:2010} and in \cite{Jung:04,Jung:07} where
the parametrization for the gluon distribution at the starting value $Q_0^2$
was found as
\be
g(k_t,x,Q_0^2)\sim R^2_0(x)k_t^2\exp(-R_0^2(x) k_t^2)~,
\label{def:gkta}
\ee
with the parameters given in \cite{Wusthof:99}. The motivation of the form given by Eq.(\ref{def:gkta})
is based on the saturation model \cite{McLerran:94}-\cite{Kovchegov:05}. Unintegated gluon distributions
were also studied in \cite{Nikolaev:2002} within the dipole model where the similar qualitative $k_t$ dependence
was shown. On the basis of these results we will try to include the contribution of gluons in soft $pp$ collisions. 
 
%%%%%%%%%%%%%%%%%%%%%%%%%%%%%%%%%%%%%%%%%%%%%%%%%%%%%%%%%%%%%%%%%%%%%%%%%%%%%%%%%%%%%%%%%%%%%%%%%%%%%%%%%%%%%%%%%%
$\bullet~${\bf Hadron production in central rapidity region}.\\
According to the Abramovskiy-Gribov-Kancheli cutting rules (AGK) \cite{AGK}, 
at mid-rapidity only Mueller-Kancheli type diagrams contribute to the inclusive spectrum
of hadrons. In our approach the function $F_\tau^{(n)}$ is calculated in such a way
that in the central region ($y=0$), when $x\simeq 0$ and $z\simeq 0$,
it becomes proportional to $n$ and satisfies AK cancellation. Thus,
\be
\rho_q(0,p_t)=\phi_q(0,p_t)\sum_{n=1}^\infty n \sigma_n(s) =
g(s/s_0)^{\Delta}\phi_q(0,p_t),
\label{def:invspq}
\ee  
where $\phi_q(x=0,p_t)$ depends only on $p_t$, $s_0=1$ Gev$^2$.
Considering the gluons from incoming protons, which may split into  $q\bar{q}$ pairs, we get an
additional contribution to the spectrum
\be
\rho_g(x=0,p_t)=\phi_g(0,p_t)\sum_{n=2}^\infty 
(n-1)\sigma_n(s)\equiv \\
\nonumber
\phi_g(0,p_t)(g(s/s_0)^{\Delta}-\sigma_{nd})~,
\label{def:invspg}
\ee
where $\Delta=0.12$, g$=$21 mb and $\sigma_{nd}$ is the cross section of exchange of any number cut-pomerons.
The quantities
\be
\sum_{n=1}^\infty n\sigma_n(s)=g(s/s_0)^{\Delta}~;~
\sum_{n=1}^\infty\sigma_n(s)=\sigma_{nd}
\label{def:gsignd}
\ee
were calculated in \cite{Ter-Mart}  within the ``quasi-eikonal'' approximation \cite{Ter-Mart}.
Assuming that one of the cut-pomerons is always stretched between valence quarks and diquarks
which are not coming from the splitting of gluons, in Eq.~(9) we excluded unity from $n$.
Finally, we can present the inclusive spectrum at $x\simeq 0$ in the following form:
\be
\rho(p_t)=\rho_q(x=0,p_t)+\rho_g(x=0,p_t)= \\
\nonumber
g(s/s_0)^{\Delta} \phi_{q}(0, p_t)+
\left(g(s/s_0^{\Delta}- \sigma_{nd}\right)
\phi_g(0, p_t)~,
\label{def:rhoagk}
\ee    
We fix these contributions from the data on the charged particles $p_t$ distribution, 
parametrizing them as follows:
\be
\phi_q(0,p_t)=A_q\exp(-b_q p_t)~\\
\nonumber
\phi_g(0,p_t)=A_g\sqrt{p_t}\exp(-b_g p_t).
\label{def:phiq}
\ee
The parameters are fixed  from the fit to the data on the $p_t$ distribution of charged particles at $y=0$:
  $A_q=0.1912\pm 0.0064;~b_q=7.24\pm 0.11$ (Gev/c)$^{-1}$ and $A_g=0.0568\pm 0.002;~ 
b_g=3.46\pm 0.02$ (GeV/c)$^{-1}$.
%%%%%%%%%%%%%%%%%%%%%%%%%%%%%%%%%%%%%%%%%%%%%%%%%%%%%%%%%%%%%%%%%%%%%%%%%%%%%%%%%%%%%%%%%%%%%%%%%%%%%%%%%%
\subsection{Hard Scattering}
%\end{center}
%$\bullet~${\bf $A+B\rightarrow h+X$}\\
As will be shown below, the approach suggested above will be able to describe many data
on the inclusive $p_t$ spectra in the central $pp$ collisions at not large values of 
$p_t\leq 2$ GeV$/$c. Therefore, at larger $p_t$ we calculate these spectra, which are due to the 
hard $pp$ collision, within the leading order of perturbative QCD (LOQCD).
According to the model of hard scattering \cite{AVEF:1974}-\cite{FF:AKK08}, the relativistic invariant
inclusive spectrum of the hard process $p+p\rightarrow h+X$ can be related to
the elastic parton-parton subprocess $i+j\rightarrow i^\prime +j^\prime$,
where $i,j$ are the partons (quarks and gluons). This spectrum can be presented in
the following general form \cite{FF}-\cite{FFF2}:
\be
\rho(x,p_t)_{hard}\equiv E\frac{d{\hat\sigma}_{hard}}{d^3p}=\sum_{i,j}\int d^2k_{it}\int d^2k_{jt}
\int_{x_i^{min}}^1dx_i\int_{x_j^{min}}^1dx_j \\
\nonumber
\int_0^1 dz
f_i(x_i,k_{it})f_j(x_j,k_{jt})
\frac{{\hat s}}{\pi}\frac{d{\sigma_{ij}}({\hat s},{\hat t})}{d{\hat t}}\frac{1}{z^2}
D_{i,j}^h(z)\delta({\hat s}+{\hat t}+{\hat u})
\label{def:rho}
\ee
where $s=(p_1+p_2)^2\simeq 2p_1\cdot p_2$, $p_1,p_2$ are the four-momenta of the colliding protons.
In the c.ms. of $pp$ (LHC facility, when ${\vec p}_1=-{\vec p}_2$)
$s=4E_1^2$, i.e., $E_1=\sqrt{s}/2$; $t=(p_1-p_h)^2\simeq -2p_1\cdot p_h=-2(E_1 E_h-{\vec p}_1{\vec p}_h)$;
$u=(p_2-p_h)^2\simeq -2p_2\cdot p_h=-2(E_1 E_h-{\vec p}_1{\vec p}_h)$; $x_{(i,j)}=2k_{{(i,j)}z}/\sqrt{s}$;
$E_{(1,2,h)}=\sqrt{m_{(1,1,h)}^2+{\vec p}_{(1,2,h)}^2}$, ${\vec p}_{(1,2,h)}$ are the three-momenta of hadrons 
$1,2,h$ respectively; $k_{(i,j)z}$ are the longitudinal momenta (relative to ${\vec p}_A$) of the partons $i$
or $j$ in the $pp$ c.m.s., $z$ is the fraction of the hadron momentum from the parton momentum,
$f_{i,j}$ is the PDF, whereas $D_{i,j}$ is the FF, ${\hat s},{\hat t},{\hat u}$ are the Mandelstam 
variables for the parton-parton elastic scattering, see the details in \cite{FF}. Calculating Eq.(\ref{def:rho})
we used the PDF, FF and $d\sigma_{ij}({\hat s},{\hat t})/d{\hat t}$ obtained within LO QCD
\cite{Mangano:2010}-\cite{FF:AKK08}.  
% $z=\mid{\vec p}_h\mid cos(\theta_{i^\prime h})\mid/\mid{\vec k}_i^\prime\mid $, 
%where $cos\theta_{i^\prime h}={\vec k}_i^\prime\cdot{\vec p}_h/(\mid{\vec k}_i^\prime\mid\mid{\vec p}_h\mid)$
%and ${\vec k}_i^\prime$ is the three-momentum of the scattered elastically parton.

%%%%%%%%%%%%%%%%%%%%%%%%%%%%%%%%%%%%%%%%%%%%%%%%%%%%%%%%%%%%%%%%%%%%%%%%%%%%%%%%%%%%%%%%%%%%%%%%%%%%%%%%%%%%%%%%%%%%
\section{Results and discussion}
In Figs.(2-5) we illustrate how our approach works for the description of the experimental data on the inclusive
spectra of pions and kaons produced in $pp$ collisions when we neglect the gluon distributions in the proton.
It allows a satisfactory description of these spectra as functions of $x$ and $p_t$, up to $p_t=1.4$ GeV$/c$ at the
initial ISR energies $\sqrt{s}=23.3-53$ GeV. The ISR experimental data presented 
in Figs.(2-5) are taken from \cite{ISR}. The quark (diquark) distribution and the FF as a function of the internal 
transverse momentum $k_t$ and $\tilde{k}_t$ are chosen in the exponential form given by Eq.(\ref{def:gkt}). We took
the same parameters $B_q$ and $B_F$ for all the initial ISR energies, namely $B_q=$4.5 GeV$/c$ for both valence and 
sea quarks of any flavour and $B_F=$2.8 GeV$/c$ for $\pi^\pm$ and $B_F=$2.4 GeV$/c$ for $K^\pm$ mesons.
Figures (2-5) show that the modified QGSM results in a self-consistent satisfactory description of the inclusive
spectra of light mesons produced in $pp$ collisions at ISR energies at different values of $x$ and $p_t$ not 
larger than 1.5 GeV$/c$. However, one cannot satisfactorily describe similar spectra at the S$p{\bar p}$S, Tevatron and LHC
energies ignoring the gluon contribution . 
   \begin{figure}[h!]
 %\rotatebox{270}
  {\epsfig{file=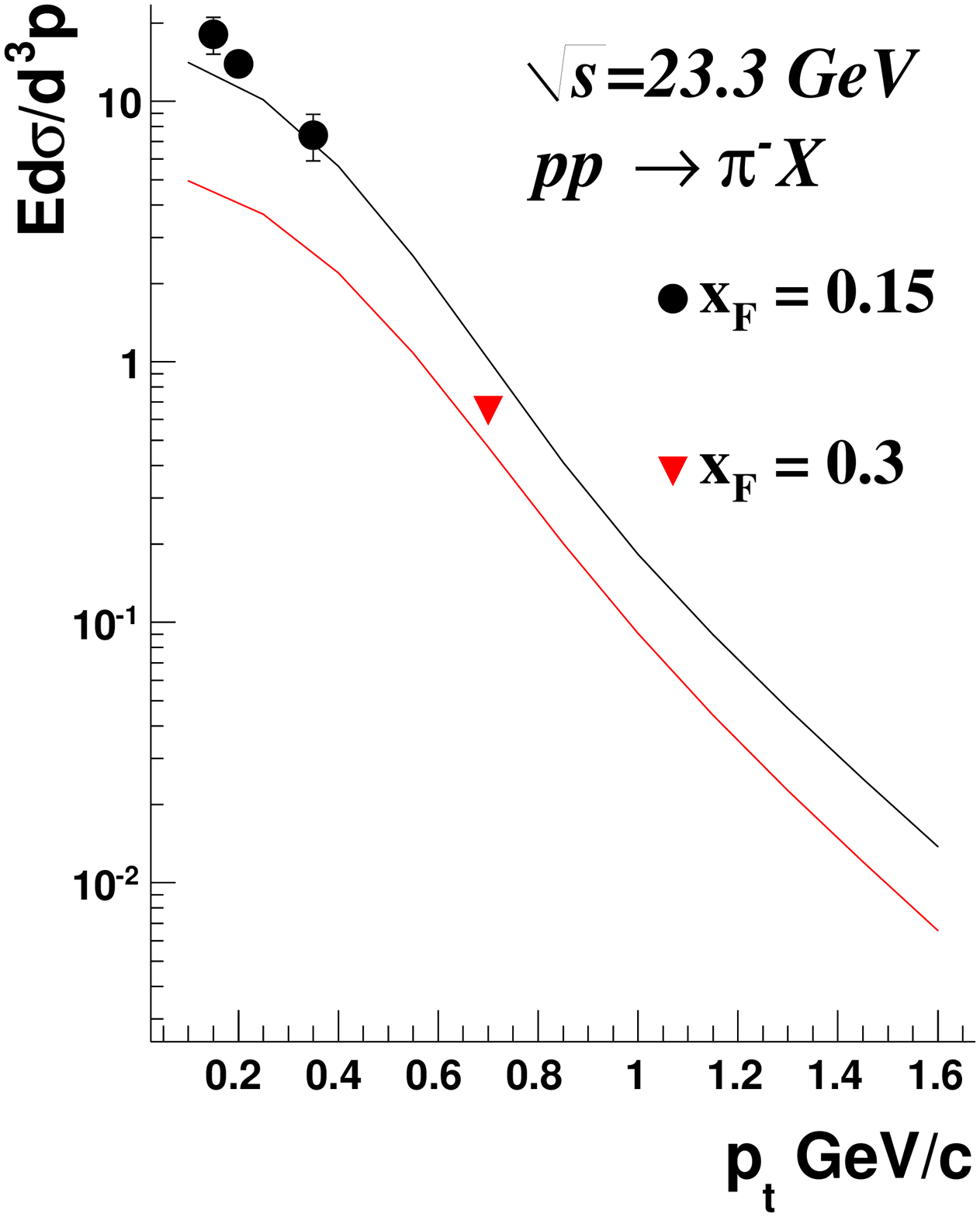,height=8cm,width=6.5cm  }}
  {\epsfig{file=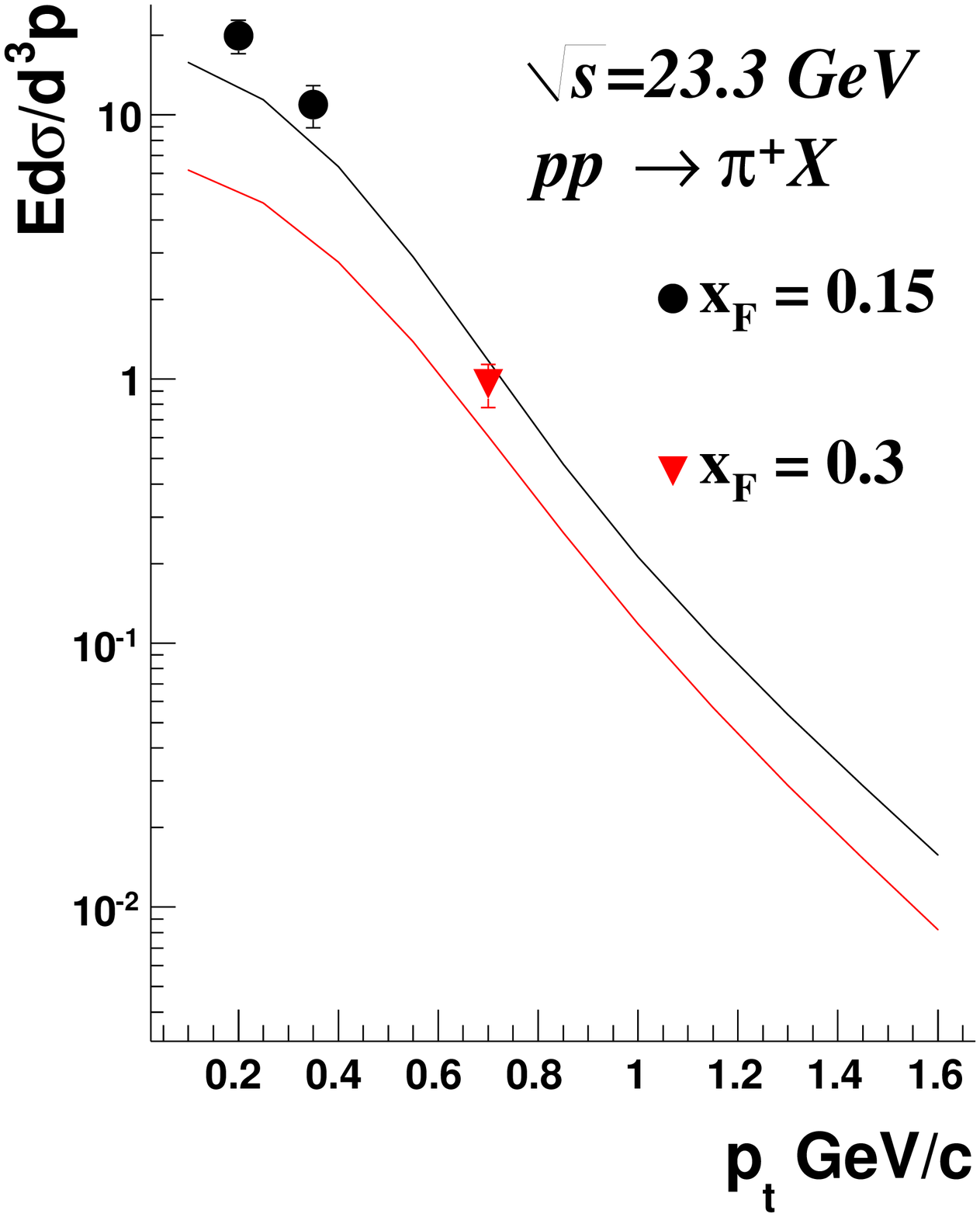,height=8cm,width=6.5cm  }}
  \caption[Fig.2]{The inclusive spectrum $Ed\sigma/d^3p$~[mbGeV$^{-2}$c$^3$]
 of $\pi^-$ mesons (left) produced in $pp$ collision at
$\sqrt{s}=23.3$ GeV; the similar spectrum but for $\pi^+$ mesons (right).}
  \end{figure}

   \begin{figure}[h!]
 %\rotatebox{270}
  {\epsfig{file=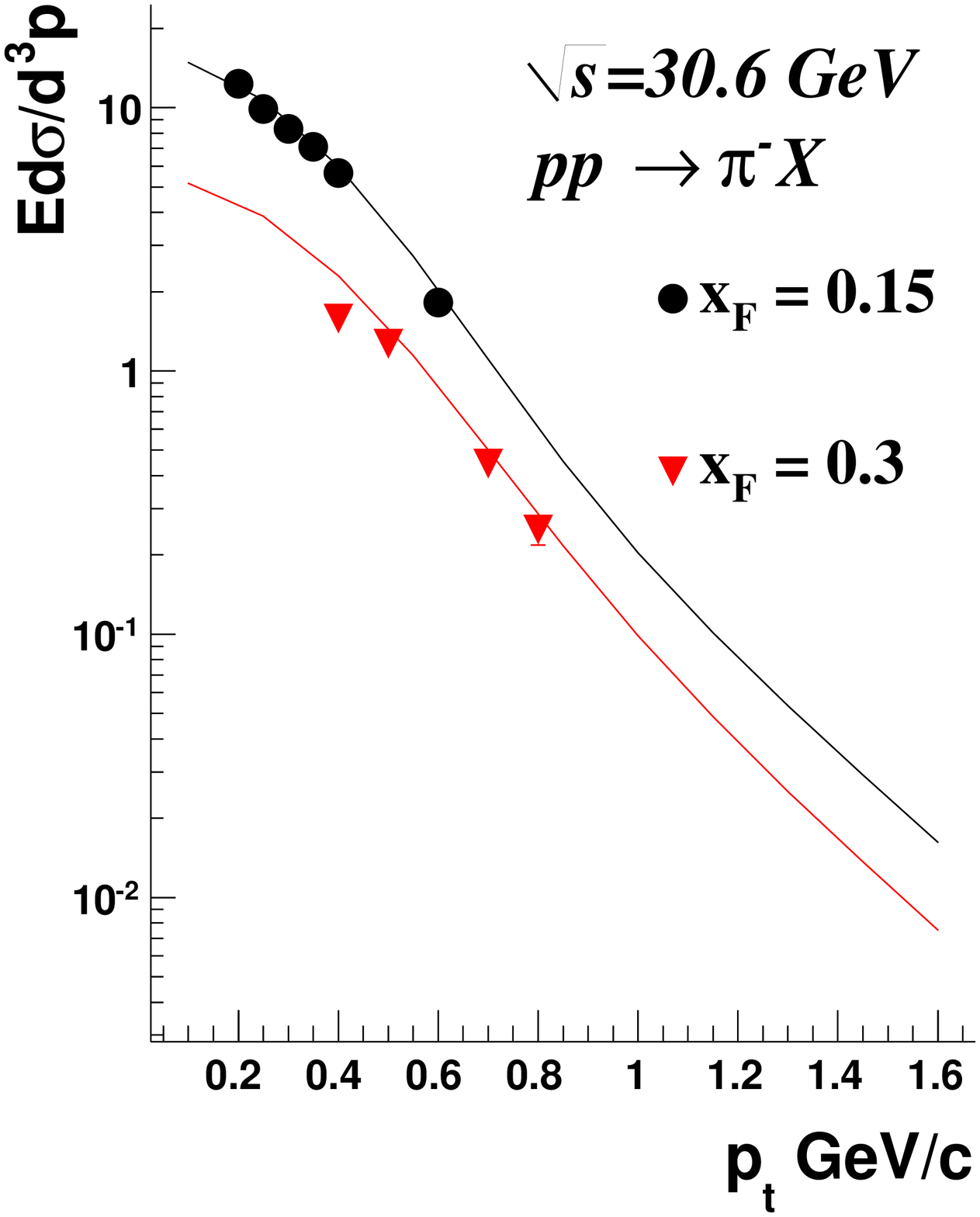,height=8cm,width=6.5cm  }}
  {\epsfig{file=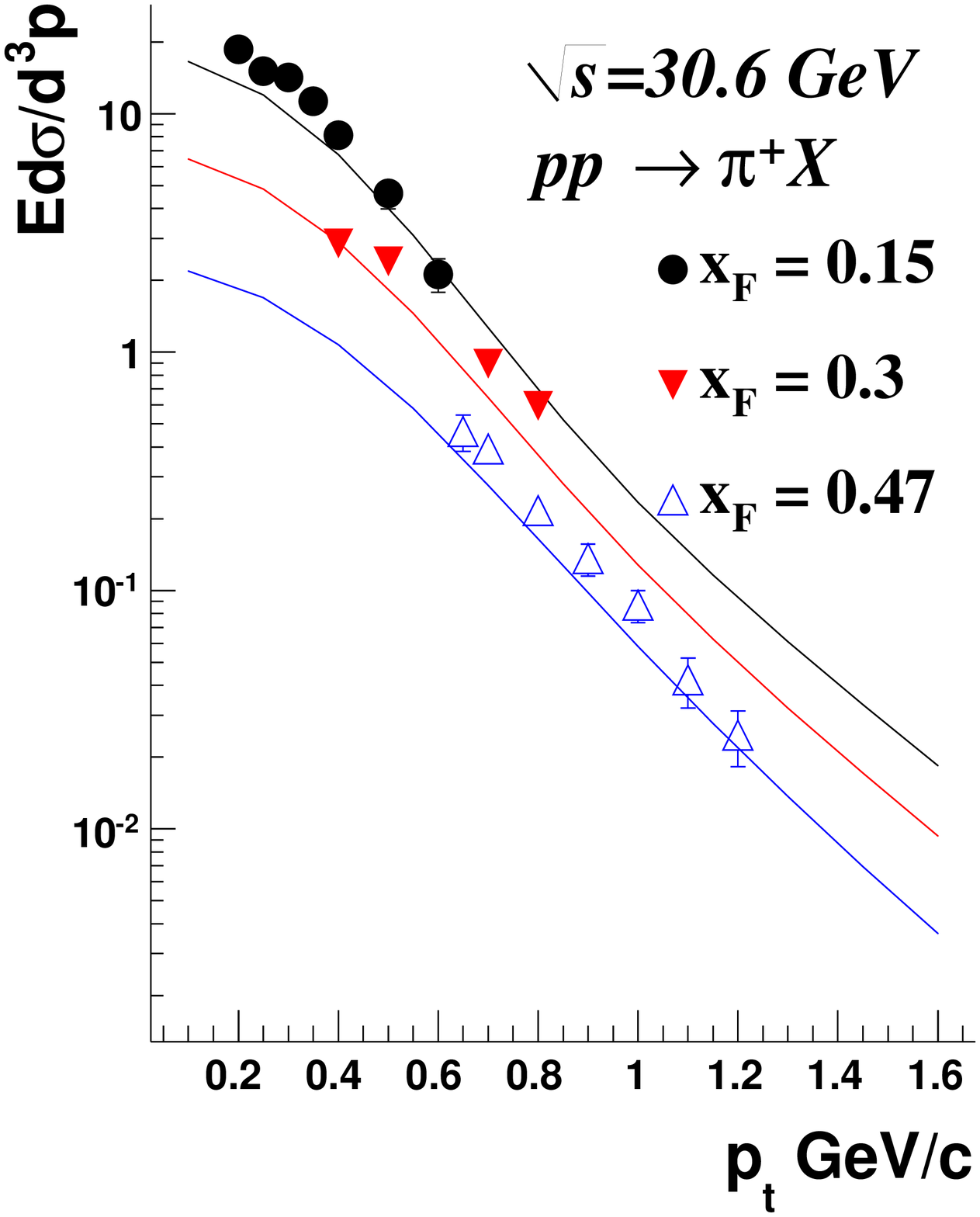,height=8cm,width=6.5cm  }}
  \caption[Fig.3]{The inclusive spectrum $Ed\sigma/d^3p$~[mbGeV$^{-2}$c$^3$]
of $\pi^-$ mesons (left) produced in $pp$ collision at
$\sqrt{s}=30.6$ GeV; the similar spectrum but for $\pi^+$ mesons (right).}
  \end{figure}

   \begin{figure}[h!]
 %\rotatebox{270}
  {\epsfig{file=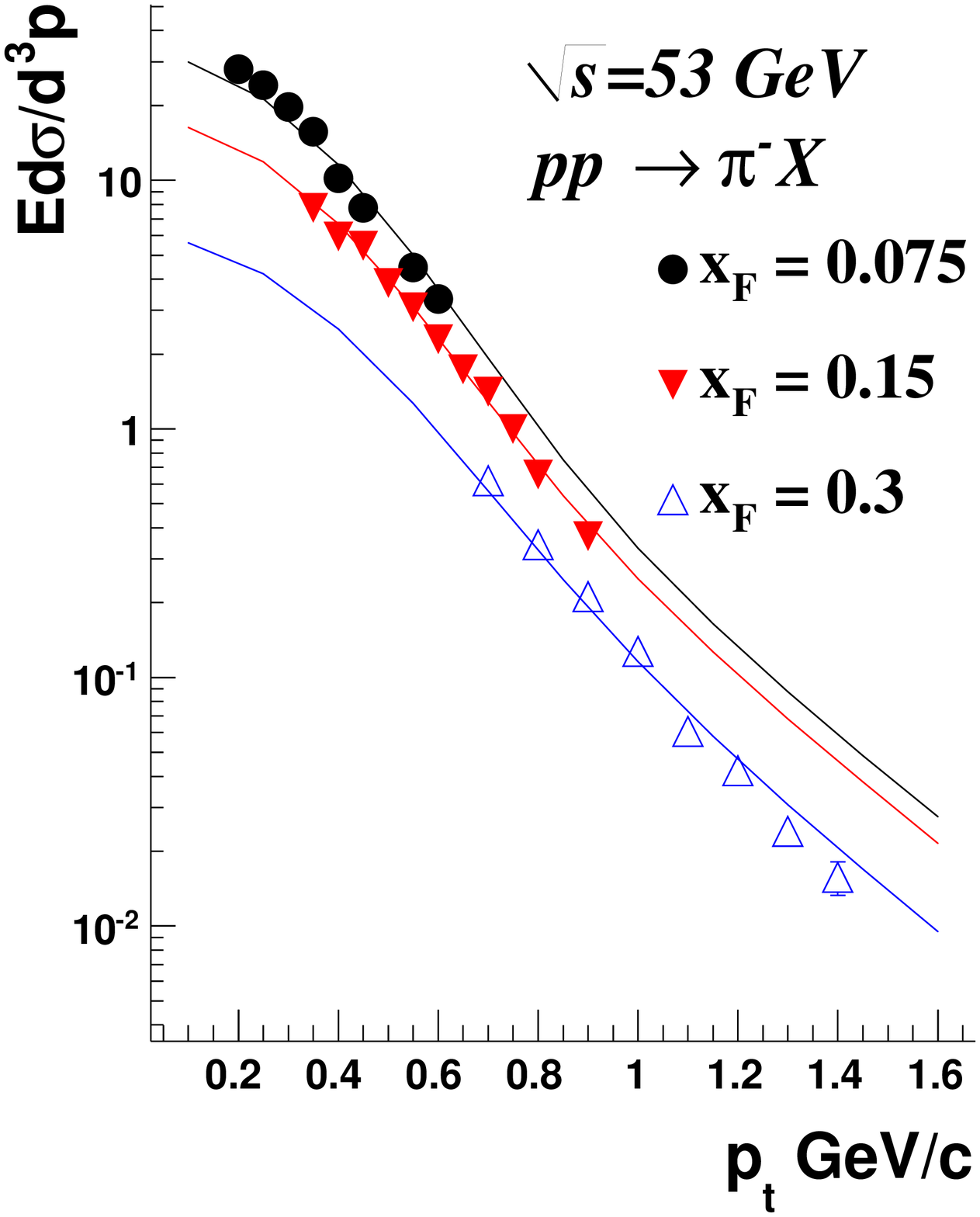,height=8cm,width=6.5cm  }}
  {\epsfig{file=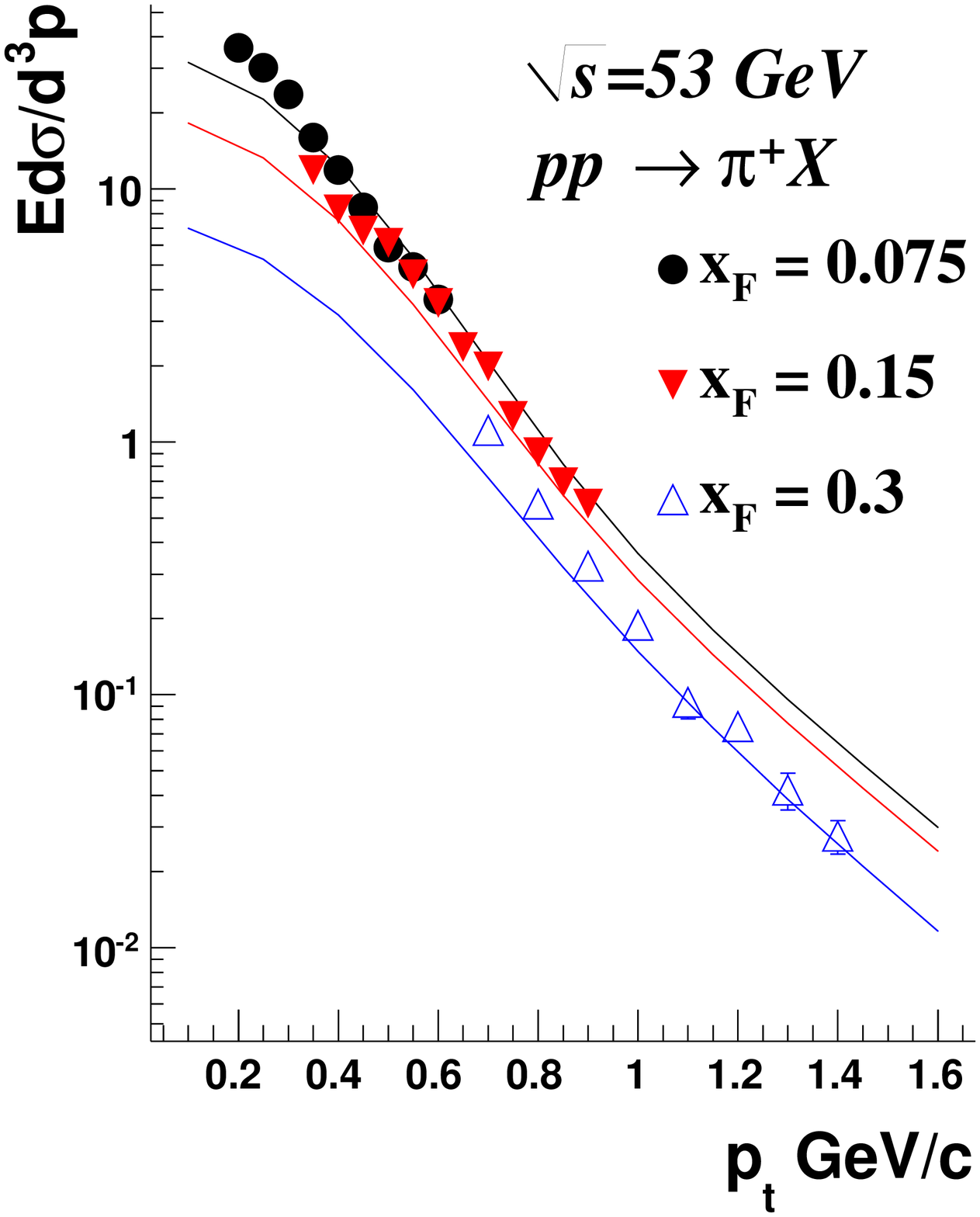,height=8cm,width=6.5cm  }}
  \caption[Fig.4]{The inclusive spectrum $Ed\sigma/d^3p$~[mbGeV$^{-2}$c$^3$]
of $\pi^-$ mesons (left) produced in $p p$ collision at
$\sqrt{s}=53$ GeV; the similar spectrum but for $\pi^+$ mesons (right).}
  \end{figure}

   \begin{figure}[h!]
 %\rotatebox{270}
  {\epsfig{file=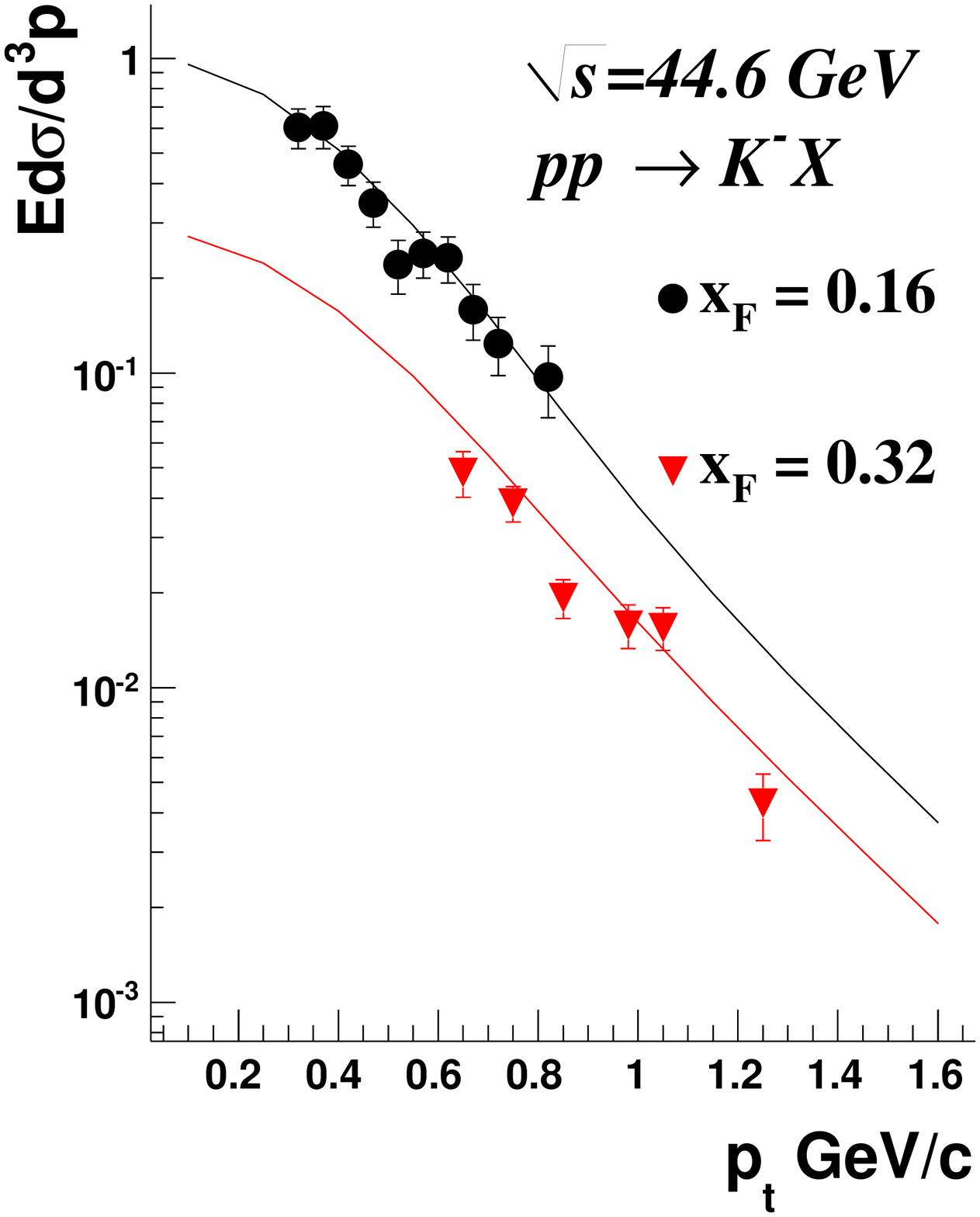,height=8cm,width=6.5cm  }}
  {\epsfig{file=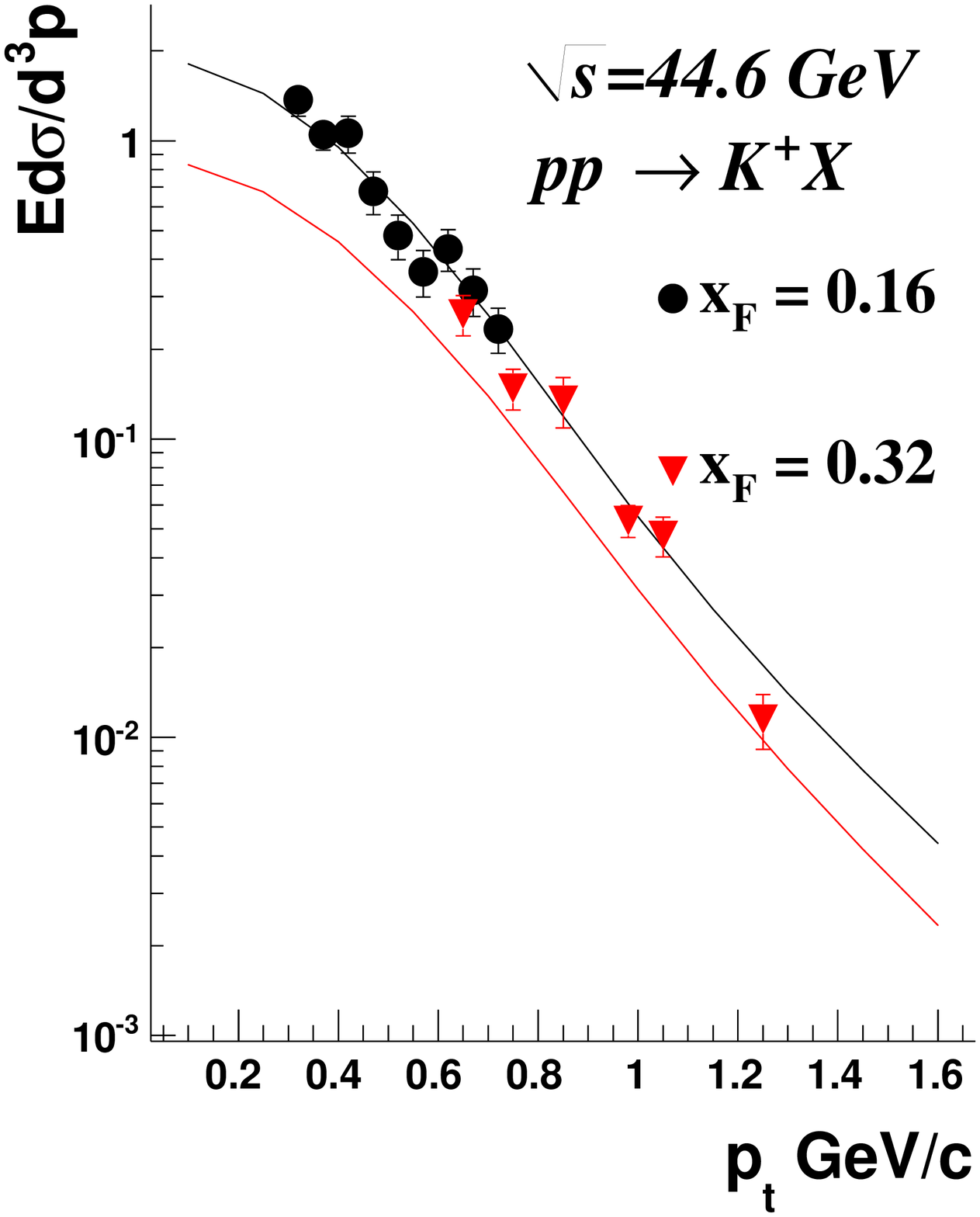,height=8cm,width=6.5cm  }}
  \caption[Fig.5]{The inclusive spectrum $Ed\sigma/d^3p$~[mbGeV$^{-2}$c$^3$]
of $K^-$ mesons (left) produced in $p p$ collision at
$\sqrt{s}=44.6$ GeV; the similar spectrum but for $K^+$ mesons (right).}
  \end{figure}

    \begin{figure}[h!]
 %\rotatebox{270}
  {\epsfig{file=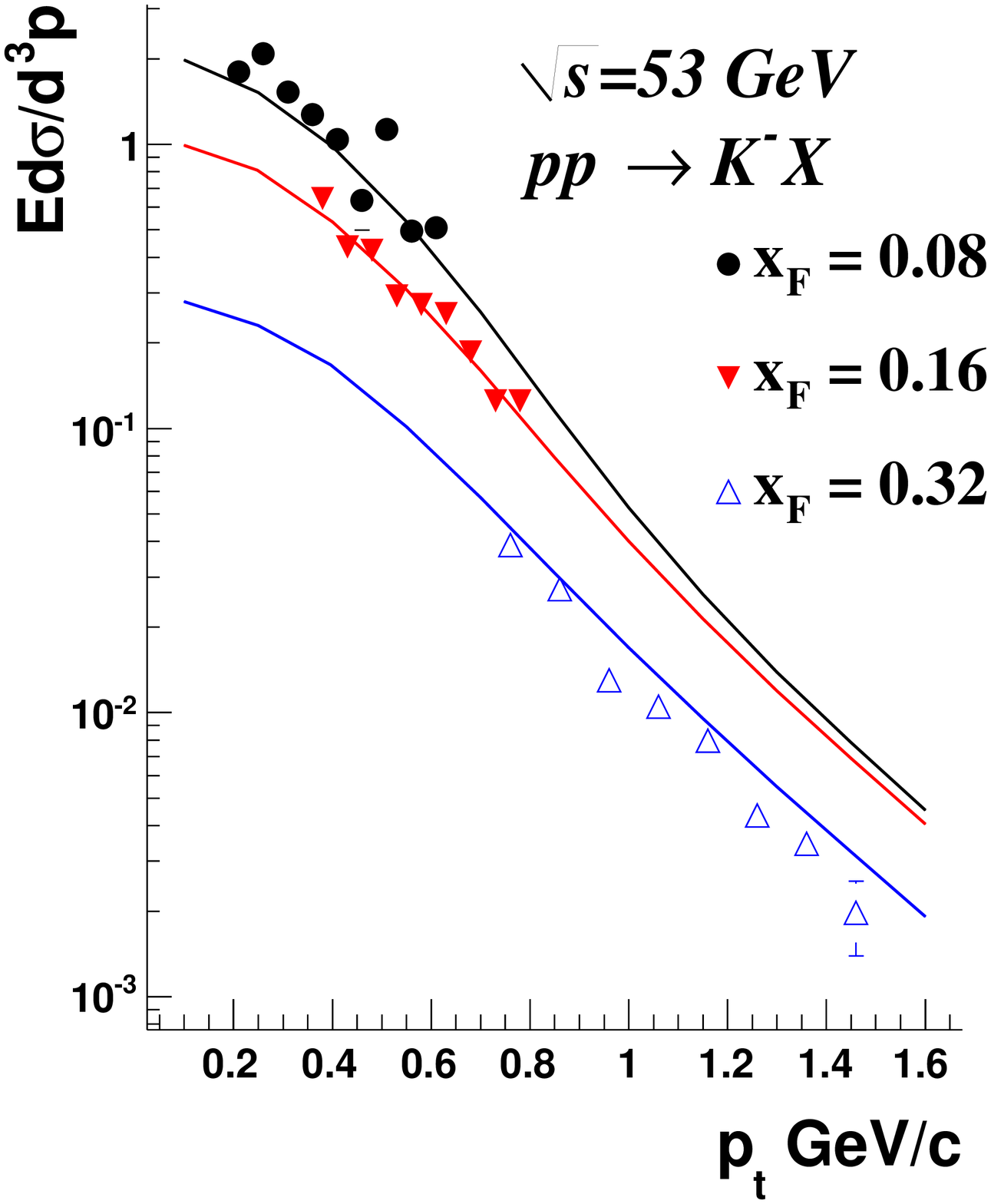,height=8cm,width=6.5cm  }}
  {\epsfig{file=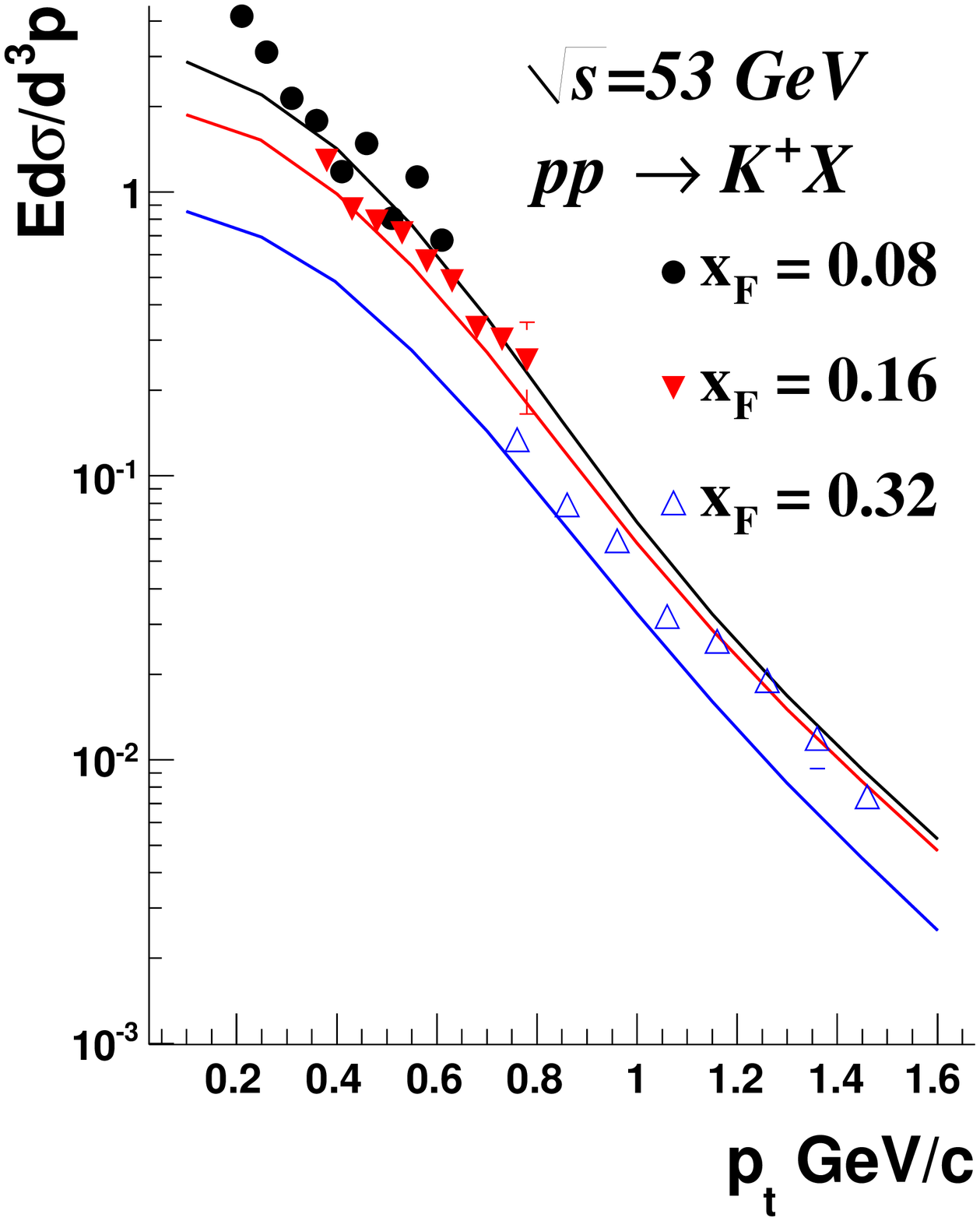,height=8cm,width=6.5cm  }}
  \caption[Fig.6]{The inclusive spectrum $Ed\sigma/d^3p$~[mbGeV$^{-2}$c$^3$]
of $K^-$ mesons (left) produced in $p p$ collision at
$\sqrt{s}=53$ GeV; the similar spectrum but for $K^+$ mesons (right).}
  \end{figure}
 
     \begin{figure}[h!]
 %\rotatebox{270}
{\epsfig{file=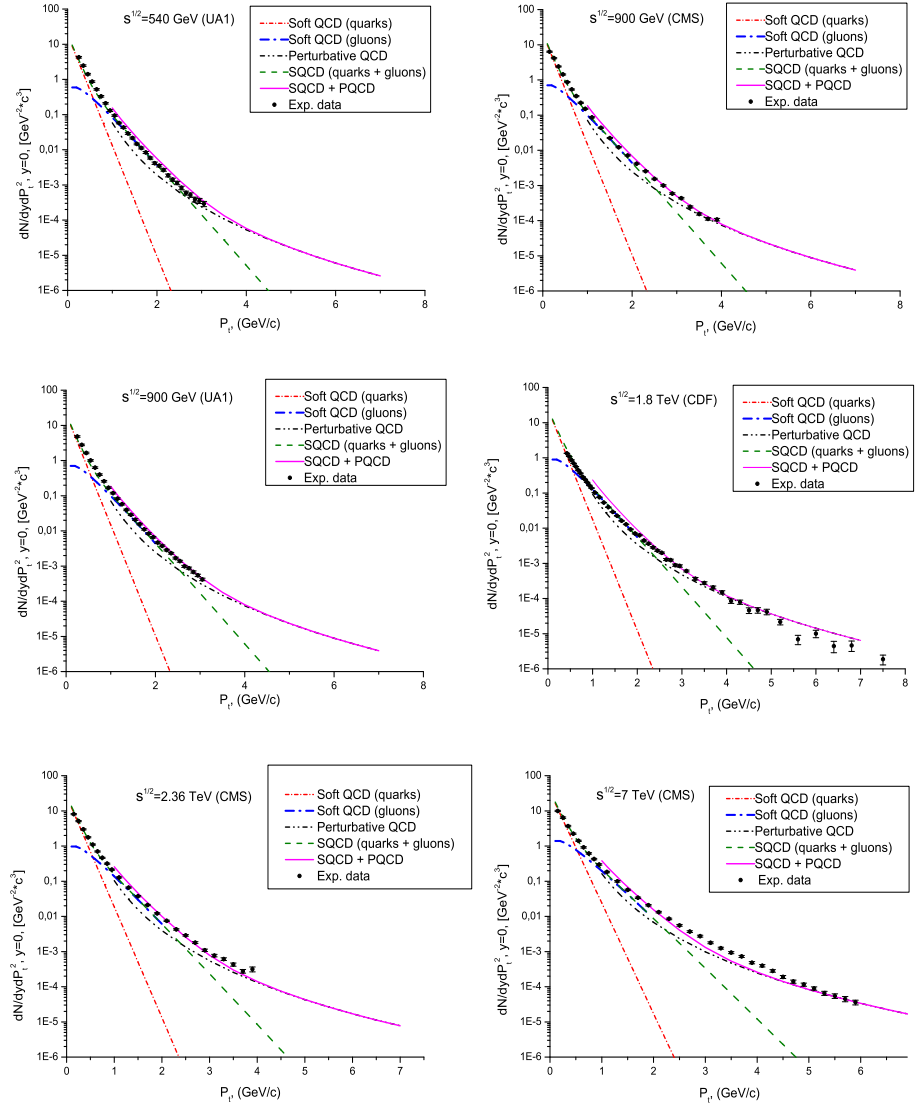,height=18cm,width=24cm  }}
%  {\epsfig{file=pt_HE_2.eps,height=9cm,width=14cm  }}
 % {\epsfig{file=K53p.eps,height=9cm,width=12cm  }}
  \caption[Fig.7]{The inclusive spectrum of the charged hadron as a function of $p_t$ (GeV$/c$)
in the central rapidity region ($y=0$) at $\sqrt{s}=540, 900$ GeV (top) and 
$\sqrt{s}=1.8, 2.36, 7$ TeV. The data are taken from \cite{UA1,CDF,CMS}.}
  \end{figure}

%     \begin{figure}[ht]
% %\rotatebox{270}
%{\epsfig{file=pt_HE.eps,height=14cm,width=16cm  }}
%%  {\epsfig{file=pt_HE_2.eps,height=9cm,width=14cm  }}
% % {\epsfig{file=K53p.eps,height=9cm,width=12cm  }}
%  \caption[Fig.2]{The inclusive spectrum of charged hadron as a function of $p_t$ (GeV$/c$)
%in the central rapidity region ($y=0$) at $\sqrt{s}=540, 900$ GeV (top) and 
%$\sqrt{s}=1.8, 2.36, 7$ TeV.}
%  \end{figure}

     \begin{figure}[h!]
 %\rotatebox{270}
  {\epsfig{file=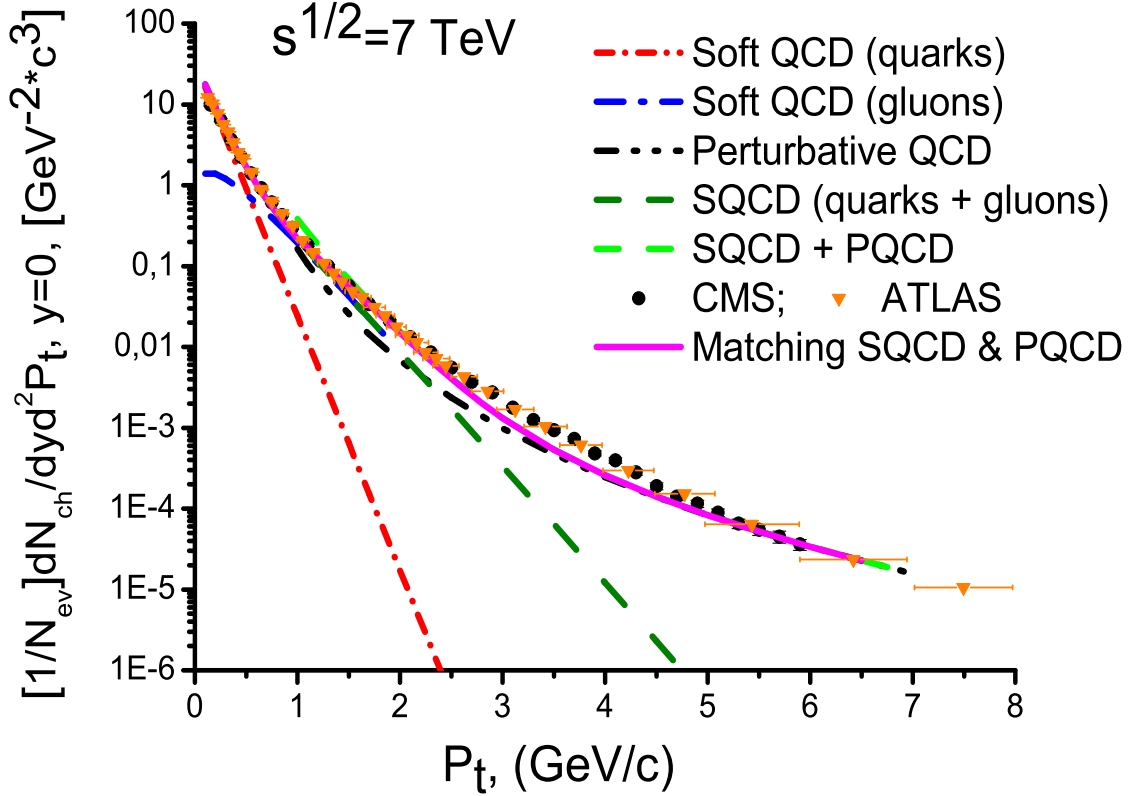,height=12cm,width=16cm  }}
 % {\epsfig{fileAndrei_G.eps=K53p.eps,height=9cm,width=12cm  }}
  \caption[Fig.8]{The inclusive spectrum of the charged hadron as a function of $p_t$ (GeV$/c$)
in the central rapidity region ($y=0$) at $\sqrt{s}=$7 TeV compared with the CMS \cite{CMS}
 and ATLAS \cite{ATLAS:2010} data.}
  \end{figure} 
 
     \begin{figure}[h!]
 %\rotatebox{270}
{\epsfig{file=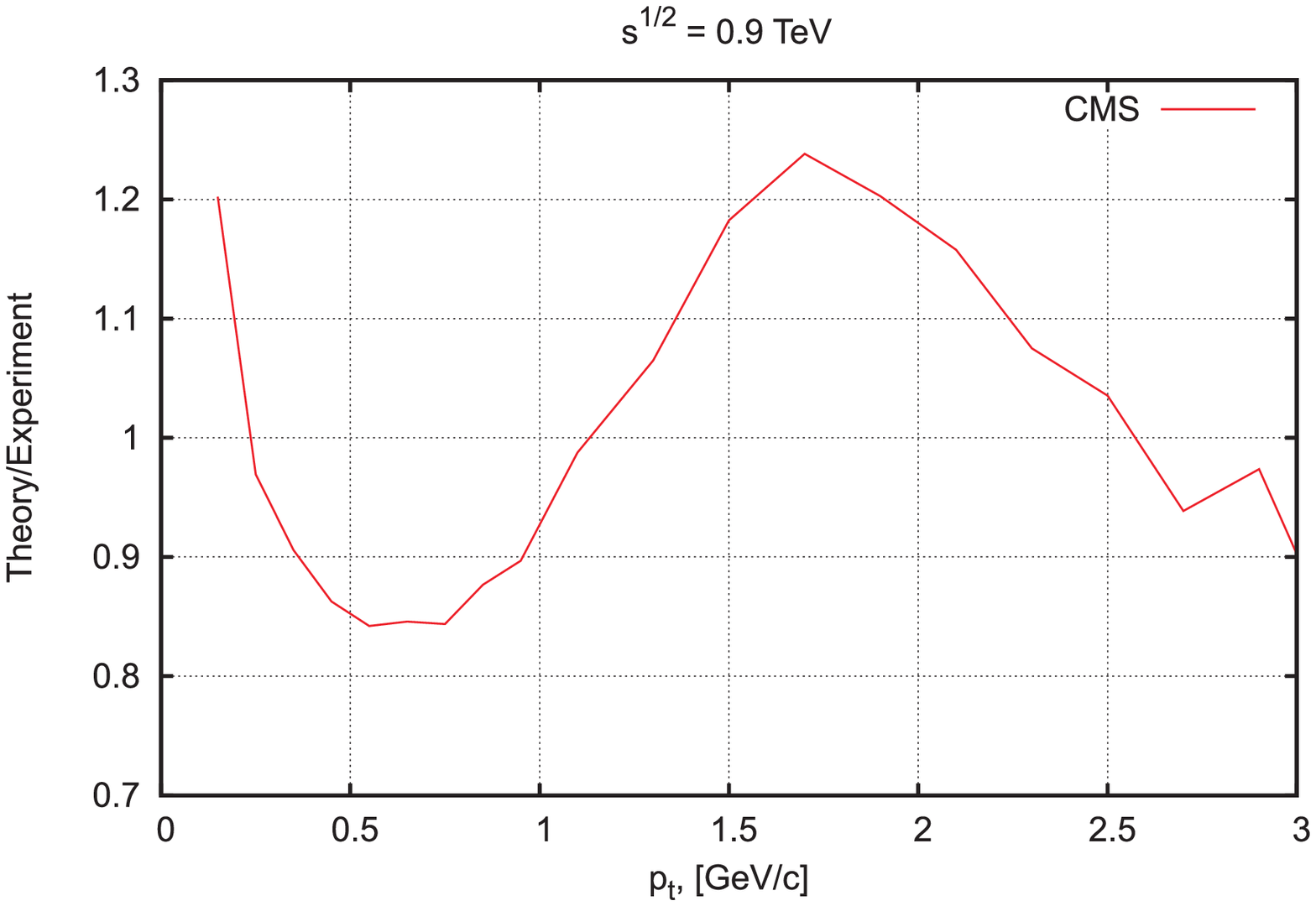,height=7cm,width=14cm  }}
  {\epsfig{file=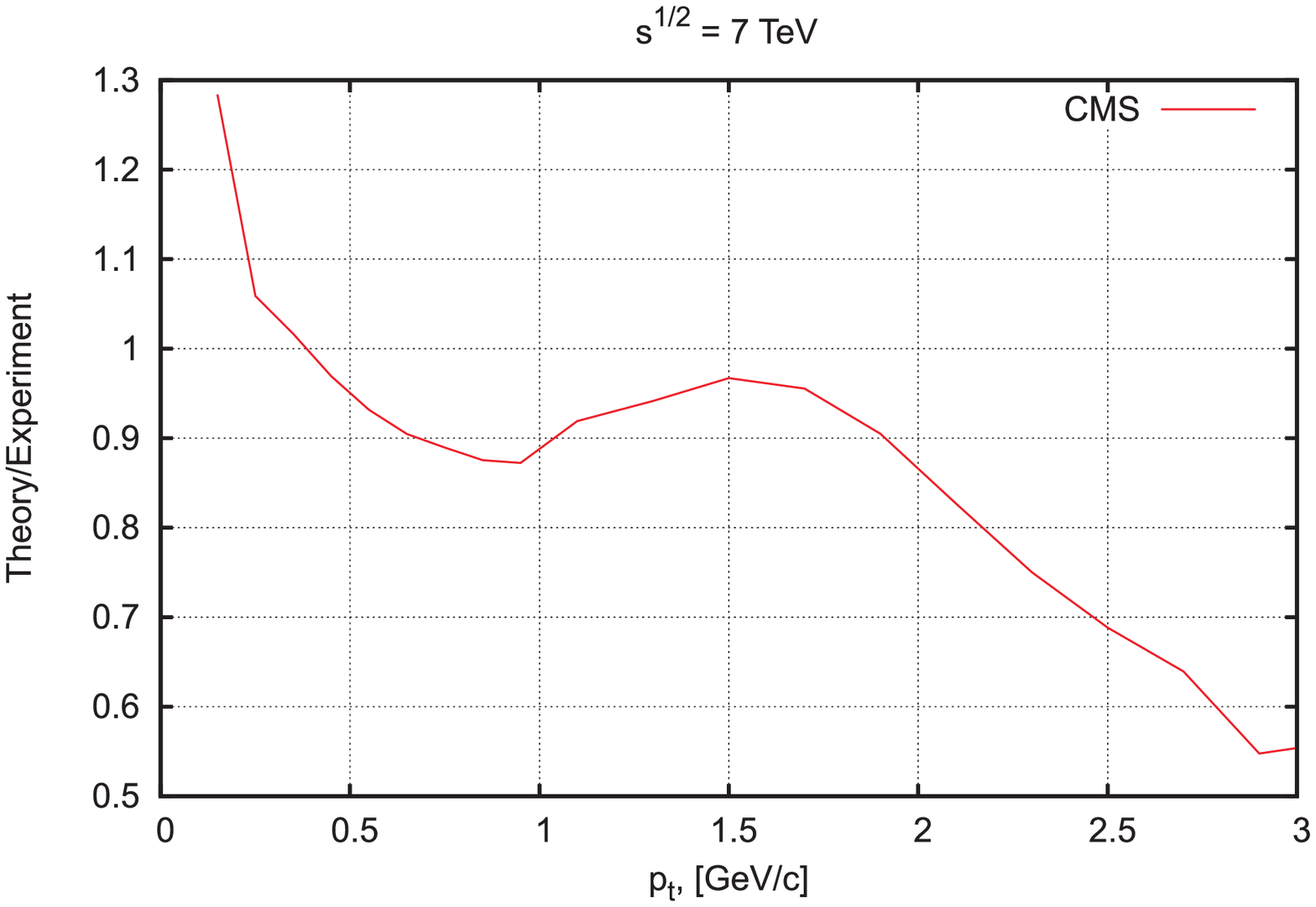,height=7cm,width=14cm  }}
 % {\epsfig{file=K53p.eps,height=9cm,width=12cm  }}
  \caption[Fig.9]{The ratio of our calculations of the inclusive $p_t$ spectrum as a function of $p_t$ (GeV$/c$)
in the central rapidity region ($y=0$) to the corresponding experimental data at $\sqrt{s}=$900 GeV (top) and 
$\sqrt{s}=$7 TeV (bottom). The data are taken from \cite{CMS}.}  
\end{figure}
The result of the fit to data using Eq.(\ref{def:rhoagk}) on the charged hadron inclusive spectra is presented 
in Figs.(7,8).
The long-dash curve corresponds to the quark contribution $\rho_q(x=0, p_t)$ given by
Eq.(\ref{def:invspq}), and the short-dash line is the gluon contribution $\rho_g(x=0, p_t)$ (Eq.17)
to the invariant yield $d^3N/dyd^2p_t$; the solid curve corresponds to the sum of both contributions, 
see Eqs.(19,20).
One can see that the conventional quark contribution $\Phi^q(y=0,p_t)$
is able to describe the data up to $p_t\leq$1 GeV$/c$, whereas the inclusion of the gluon contribution allows 
us to extend the range of good description up to 2 GeV$/c$. At larger values of $p_t$ the contribution of hard
processes is not negligible and one has to take them into account on the basis of the perturbative QCD.
Figure 7 also shows that, according to the experimental data, the shape of the inclusive $p_t$ spectrum at $y=0$
slowly changes as the initial energy increases. Its  $p_t$ dependence appears harder as the energy increases.
It is probably due to the jet production which occurs even at not large values of $p_t$ when $\sqrt{s}$ increases.
In Fig.8 the same results as in Fig.7(right, bottom) except the solid line are presented. The solid line in Fig.8
corresponds to the matching of the calculations obtained within the soft QCD (SQCD) and perturbative QCD (PQCD).
The SQCD means our calculations within the QGSM including the gluons in the proton, see Eq.(19)
In Fig.9 the ratio of our calculations to the experimental data \cite{CMS} at $\sqrt{s}=$900 GeV (top)
and $\sqrt{s}=$7 TeV are presented. It is shown that the discrepancy between theoretical results and the 
experimental data is less than 30 percent at $p_t\leq 2.$GeV$/$c.

The effective inclusion of the unintegrated distributions of gluons in the proton \cite{Jung:04,Jung:07}
allows us to extend the satisfactory description of the experimental data on
the inclusive spectra of charged hadrons at $y=0$ up to $p_t\sim 2 GeV/c$. At higher values of
the transverse momentum we include hard $pp$ collision calculated within LO QCD \cite{AVEF:1974}-\cite{FF:AKK08}.
Parton interactions should be considered for describing the experimental data. These results are presented in
Figs.(7,8) at $p_t>$ 2GeV$/$c.

\section{Conclusion}
Our study has shown that soft QCD or the modified QGSM including both the longitudinal and transverse motion of partons
in the proton is able to describe rather satisfactorily the experimental data on inclusive spectra of light
hadrons like pions and kaons produced in $pp$ collisions at the ISR energies and at 
not large values of the transverse momenta $p_t\leq 1.4-1.5 GeV/c$. These calculations were made in the self-consistent way,
i.e., the parameters entering into the quark distribution and the FF, as a function of the internal transverse momentum, 
are the same at all the initial energies $\sqrt{s}$ up to about a few hundred GeV.
%When the initial energy increases we 
%need to change these values to describe the experimental data. 
The conventional QGSM or DPM model does not include the distribution of gluons in the proton. However,   
as is well known, at large transverse momenta $p_t$ of hadrons the gluons in the proton play a very important role in 
description of the
experimental data. Therefore, one can assume that the contribution of the gluon distribution in the proton to the inclusive 
spectrum 
of produced hadrons slowly appears when $p_t$ increases and it will be  sizable at high values of $p_t $. This assumption is 
also confirmed by the increase of the unintegrated gluon distribution in the proton at $x\sim 0$ as a function of the internal 
transverse 
momentum $k_t$ when $k_t$ grows \cite{Ryskin:2010,Kochelev:1998,Jung:04}.

Therefore, to illustrate this hypothesis we fit the experimental data on the inclusive spectra of charged particles
produced in the central $pp$ collisions at energies larger than the ISR starting from 500 GeV up to 7 TeV by the sum of
the quark contribution $\rho_q$ given by Eq.(\ref{def:invspq}) and the gluon contribution $\rho_g$ (see Eq.(\ref{def:invspg})).
 The parameters of this fit do not depend on the initial energy in that energy interval. This fit shows that the inclusion of
only the quark contribution  $\rho_q$ allows us to describe the data at $y=0 $ up to $p_t\sim 1$ GeV$/c$, while the inclusion
of the gluon contribution $\rho_g$ results in a satisfactory description of the data up to $ p_t\simeq $2GeV$/c$.
The inclusion of the hard $pp$ collision within LO QCD allows us to describe rather satisfactorily the data in the wide 
region of $p_t$.
%At higher values of the transverse momentum the hard parton interactions should be considered for describing the experimental 
%data. 
%%%%%%%%%%%%%%%%%%%%%%%%%%%%%%%%%%%%%%%%%%%%%%%%%%%%%%%%%%%%%%%%%%%%%%%%%%%%%%%%%%%%%%%%%%%%
\vspace{0.1cm}

{\bf Acknowledgements}\\
The authors are very grateful to A.Bakulev, V.Cavazini, A.Dorokhov, A.V.Efremov, F.Francavilla, 
C. Gwenlan, H.Jung, V.Kim, B.Kniehl, N.Kochelev, E.A.Kuraev, L.N.Lipatov, T.Lomtadze, M.Mangano, C.Merino, 
S.V.Mikhailov, E.Nurse, F.Palla, E.Pilkington, A.F.Pikelner, C.Royon, M.G.Ryskin, E.Sarkisyan-Grinbaum, 
O.V.Teryaev, Yu.Shabelski and V.V.Uzhinskiy
for very useful discussions and comments. This work was supported in part by the Russian Foundation for Basic 
Research, project No:11-02-01538-a.

\section{Appendix}

$\bullet~${\bf Quark distributions in proton within QGSM}\\
Let us present the quark distributions in a proton obtained within the Regge theory in 
Refs.\cite{kaid1,kaid2,Kaid1}.
The distributions for valence $u$ and $d$ quarks in the $n$ chain (see Fig.1) read
\be
f_{u_v}^{(n)}(x)=C_u x^{-\alpha_R(0)}(1-x)^{\alpha_R(0)-2\alpha_N(0)+n-1}~,~
f_{d_v}^{(n)}(x)=f_{u_v}^{(n)}(x)\cdot(1-x)~.
\label{def:fu}
\ee
There are the following relations for the sea $u$ and $d$ quarks:
\be
f_{u_{sea}}^{(n)}(x)=f_{{\bar u}_{sea}}^{(n)}(x)=f_{u_v}^{(n)}(x)~;~
f_{d_{sea}}^{(n)}(x)=f_{{\bar d}_{sea}}^{(n)}(x)=f_{d_v}^{(n)}(x)~.
\label{def:fsea}
\ee
For the charmed quarks $c{\bar c}$ in a proton we have \cite{LAS}
\be
f_{c{\bar c}}^{(n)}(x)=C_{c{\bar c}} x^{\alpha_\psi(0)}(1-x)^{2(\alpha_R(0)-\alpha_N(0))-\alpha_\psi(0)+n-1}
\label{def:fc}
\ee
The distributions of the diquarks $ud$ and $uu$ in a proton read
\be
f_{ud}^{(n)}(x)=C_{ud }x^{\alpha_R(0)-2\alpha_N(0)}(1-x)^{-\alpha_R(0)+n-1}~;~
f_{uu}^{(n)}(x)=f_{ud}^{(n)}(x)\cdot (1-x)~.
\label{def:fud}
\ee
Here $\alpha_R(0),\alpha_N(0)$ and $\alpha_\psi(0)$ are the intercepts of the Reggeon, nucleon
and $\psi$ Regge trajectories. As is well known (see, for example, \cite{Kaid1}), $\alpha_R(0)=0.5$ and
$\alpha_N(0)\simeq -0.5$.
The value for the intercept of the $\psi$ Regge trajectory is known not so well because Regge trajectories
of heavy mesons can be nonlinear as functions of the transfer $t$. For the linear $\psi$ Regge trajectory
$\alpha_\psi(0)=-2.18$, whereas for the nonlinear $\psi(t)$ the intercept value can be about zero, $\alpha_\psi(0)=0$
\cite{Piskunova}. The coefficients $C_i$ in Eqs.(\ref{def:fu}-\ref{def:fud}) are determined by the
normalization condition 
\be
\int_0^1 f_i^{(n)}(x)dx~=~1~.
\label{def:normfi}
\ee
$\bullet$~{\bf Fragmentation functions of quarks (diquarks) to $\pi$ mesons
 within QGSM}\\
The fragmentation functions of quarks (diquarks) to $\pi$ mesons 
$G_{q(qq)\rightarrow\pi}(z)=z D_{q(qq)\rightarrow\pi}$ have the following forms 
\cite{Kaid1,Shabelsky,BFL:2000}:
\be
G_{u\rightarrow\pi^+}(z)=a_0(1-z)^{\alpha_R(0)+\lambda}~;~G_{u\rightarrow\pi^-}(z)=
(1-z)G_{u\rightarrow\pi^+}~,
\ee
\be
G_{d\rightarrow\pi^+}(z)=G_{u\rightarrow\pi^-}(z)~,~G_{d\rightarrow\pi^-}(z)=
G_{u\rightarrow\pi^+}(z)~,
\ee
\be
G_{uu\rightarrow\pi^+}(z)=a_0(1-z)^{\alpha_R(0)-2{\tilde\alpha}_B(0)+\lambda}~;~
G_{uu\rightarrow\pi^-}(z)=(1-z)G_{uu\rightarrow\pi^+}(z)~,
\ee
\be
G_{ud\rightarrow\pi^+}(z)=G_{ud\rightarrow\pi^-}(z)=
a_0\left(1+(1-z)^2\right)(1-z)^{\alpha_R(0)-2{\tilde\alpha}_B(0)+\lambda}
\ee
and
\be
G_{dd\rightarrow\pi^-}(z)=G_{uu\rightarrow\pi^+}(z)~;~
G_{dd\rightarrow\pi^+}(z)=G_{uu\rightarrow\pi^-}(z)~.
\ee
In the above equations $\alpha_R(0)=0.5,{\tilde\alpha}_B(0)=-0.5, a_0=0.65$ and
$\lambda=2\alpha_R^\prime(0)<p_t^2>\simeq 0.5$, $\alpha_R^\prime(0)$ and $<p_t^2>$
being the slope of the Regge trajectory and the mean value of the Tahoe transverse hadron
momentum squared.\\
$\bullet$~{\bf FF for $K^\pm$ mesons within QGSM}\\
The FFs for $K^\pm$ mesons are \cite{ABK_OP:1985}
\be
G_{u\rightarrow K^+}(z)= G_{d\rightarrow K^0}(z)=a_K(1-z)^{-\alpha_\phi(0)+\lambda}
(1+a_{1K}z)~,
\label{def:uKpl}
\ee
\be
G_{u\rightarrow K^-}(z)= G_{d\rightarrow{\bar K}^0}(z)=G_{d\rightarrow {\bar K}^+}(z)=
G_{u\rightarrow K^0}(z)=G_{d\rightarrow {\bar K}^-}(z)=\\
\nonumber
G_{d\rightarrow {\bar K}^0}(z
a_K(1-z)^{-\alpha_\phi(0)+\lambda+1}~,
\label{def:uKmn}
\ee
\be
G_{{\bar u}\rightarrow K^-}(z)=G_{u\rightarrow K^+}(z)~,~
G_{{\bar u}\rightarrow K^+}(z)=G_{u\rightarrow K^-}(z)~,\\
\nonumber
G_{{\bar d}\rightarrow K^-}(z)=G_{d\rightarrow K^+}(z)~,~
G_{{\bar d}\rightarrow K^+}(z)=G_{d\rightarrow K^-}(z)~,~
\label{def:buKpm}
\ee
\be
G_{{\bar s}\rightarrow K^0}(z)=G_{{\bar s}\rightarrow {\bar K}^+}(z)=
G_{s\rightarrow K^-}(z)=G_{s\rightarrow {\bar K}^0}(z)=\\
\nonumber
b_Kz^{1-\alpha_\phi(0)}(1-z)^{-\alpha_R(0)+\lambda}+
a_K(1-z)^{-\alpha_R(0)+\lambda+2(1-\alpha_\phi(0))}
\label{def:bsKpl}
\ee
\be
G_{{\bar s}\rightarrow K^-}(z)=G_{{\bar s}\rightarrow {\bar K}^0}(z)=
G_{s\rightarrow K^+}(z)=G_{s\rightarrow {\bar K}^0}(z)=
%\\
%\nonumber
a_K(1-z)^{-\alpha_R(0)+\lambda+2(1-\alpha_\phi(0))}
\label{def:bsKmn}
\ee
\be
G_{uu\rightarrow K^+}(z)=a_K(1-z)^{2\alpha_R(0)-\alpha_\phi(0)
-2\alpha_N(0)+\lambda}(1+a_{2K}z)~,
\label{def:uuKp}
\ee
\be
G_{ud\rightarrow K^+}(z)=\frac{a_K}{2}(1-z)^{2\alpha_R(0)-\alpha_\phi(0)
-2\alpha_N(0)+\lambda}(1+a_{2K}z+(1-z)^2)~,
\label{def:udKp}
\ee
\be
G_{uu\rightarrow K^-}(z)=G_{uu\rightarrow K^0}(z)=G_{uu\rightarrow{\bar K}^0}(z)
a_K(1-z)^{-\alpha_\phi(0)-2\alpha_N(0)+\lambda+2}~,
\label{def:uuKm}
\ee
\be
G_{ud\rightarrow K^-}(z)=G_{ud\rightarrow{\bar K}^0}(z)=
\frac{a_K}{2}(1-z)^{-\alpha_\phi(0)
-2\alpha_N(0)+\lambda+2}(1+(1-z)^2)~,
\label{def:udKm}
\ee
\be
G_{ud\rightarrow K^0}(z)=
\frac{a_K}{2}(1-z)^{-\alpha_\phi(0)
-2\alpha_N(0)+\lambda+2}(2+a_{2K}z)~,
\label{def:udKz}
\ee
Here $\lambda=2\alpha^\prime\bar {p_t^2}\simeq 0.5,~\alpha_R(0)=0.5,~ \alpha_N(0)\simeq -0.5,~
a_{1K}=2,~a_{2K}=5,~b_K=0.4$.
$\bullet$~{\bf FF for $p$ and ${\bar p}$ within QGSM}\\
Now, the FFs for protons $p$ and antiprotons $\bar p$ are \cite{Shabel:2006}
\be
G_{u\rightarrow p}(z)=a_{\bar p}(1-z)^{\alpha_R(0)-2\alpha_N(0)+\lambda}~;~G_{u\rightarrow\pi^-}(z)=
(1-z)G_{u\rightarrow\pi^+}~,
\label{def:up}
\ee
\be
G_{d\rightarrow p}(z)=\frac{a_{\bar p}}{3}(1-z)^{\alpha_R(0)-2\alpha_N(0)+\lambda}
(3-z)~;
\label{def:dp}
\ee
\be
G_{uu\rightarrow p}(z)=G_{1uu\rightarrow p}(z)+G_{2uu\rightarrow p}(z)~,
\label{def:uup}
\ee
\be
G_{1uu\rightarrow p}(z)=a_pz^{2(\alpha_R(0)-\alpha_N(0))}(1-z)^{-\alpha_R(0)+\lambda}
(3-2z)~,
\label{def:uuone}
\ee
\be
G_{2uu\rightarrow p}(z)=a_p(1-z)^{-\alpha_R(0)+\lambda+4(1-\alpha_N(0))}~,
\label{def:uuone}
\ee
\be
G_{1ud\rightarrow p}(z)=a_pz^{2(\alpha_R(0)-\alpha_N(0))}(1-z)^{-\alpha_R(0)+\lambda}~,
\label{def:udone}
\ee
\be
G_{2ud\rightarrow p}(z)=G_{2uu\rightarrow p}(z)~,~G_{u\rightarrow {\bar p}}(z)=
G_{{\bar u}\rightarrow p}(z)=G_{d\rightarrow {\bar p}}(z)=
\label{def:udsec}
\ee
\be
G_{{\bar d}\rightarrow p}(z)=G_{{\bar u}\rightarrow p}(z)=
a_{\bar p}(1-z)^{-\alpha_R(0)+2(1-\alpha_N(0))+\lambda}
\ee
\be
G_{uu\rightarrow{\bar p}}(z)=G_{ud\rightarrow{\bar p}}(z)=
a_{\bar p}(1-z)^{\alpha_R(0)-2\alpha_N(0)+2(1-\alpha_N(0))+\lambda}
\label{def:uubp}
\ee
\be
G_{{\bar u}\rightarrow{\bar p}}(z)=G_{{\bar d}\rightarrow{\bar p}}(z)=
G_{u\rightarrow p}(z)~.
\label{def:bubp}
\ee
Here $\alpha_p(0)=0.9$ and $\alpha_{\bar p}(0)=0.07$.\\
%\newpage
%%%%%%%%%%%%%%%%%%%%%%%%%%%%%%%%%%%%%%%%%%%%%%%%%%%%%%%%%%%%%%%%%%%%%%%%%%%%%%%%%%%%%%%%%%%%%%%%%%%%%%%%%
$\bullet~${\bf Hard scattering }\\
     \begin{figure}[h!!]
 %\rotatebox{270}
  {\epsfig{file=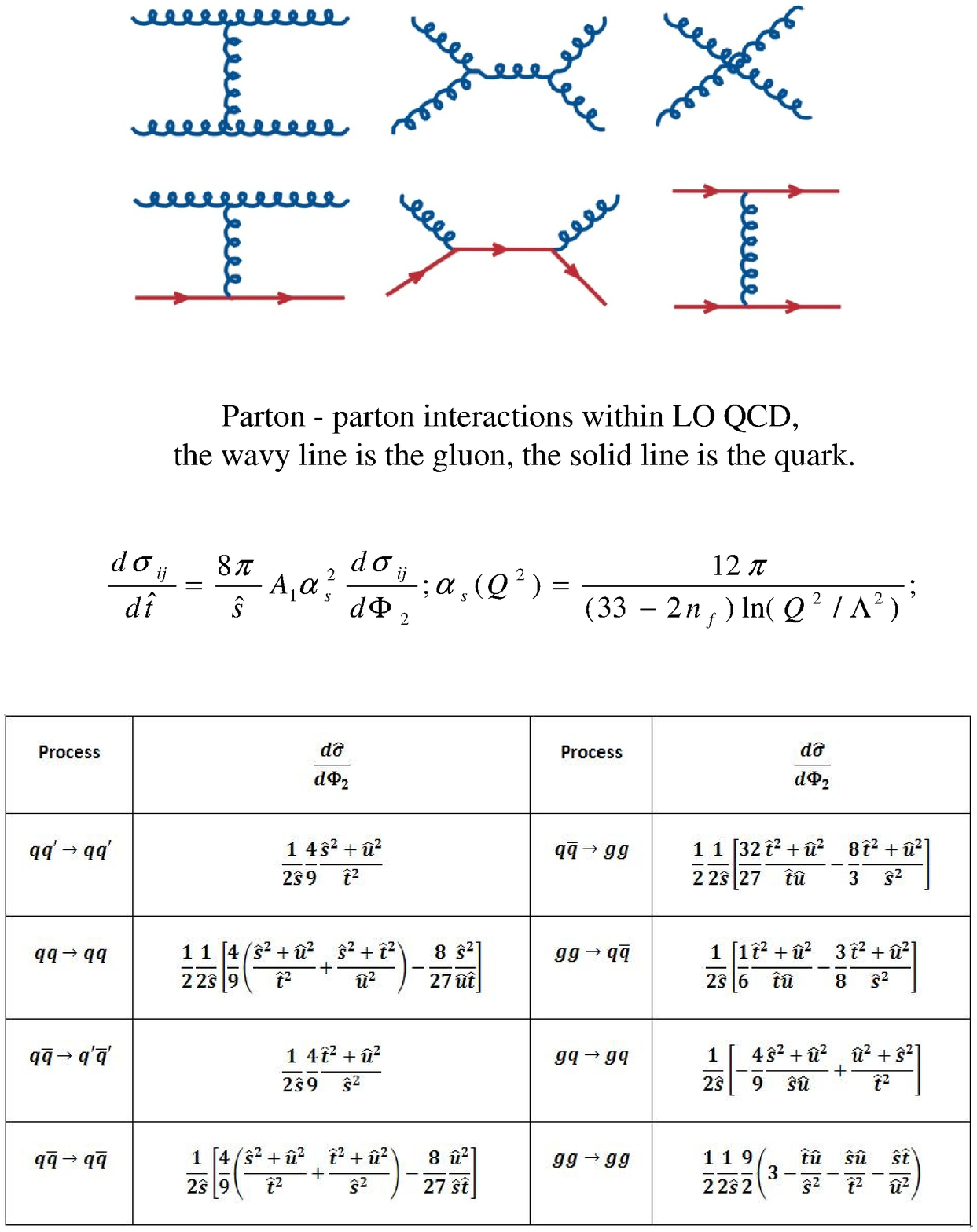,height=12cm,width=14cm  }}
 % {\epsfig{fileAndrei_G.eps=K53p.eps,height=9cm,width=12cm  }}
  \caption[Fig.9]{The elastic parton-parton scattering within the LO QCD, the wavy line 
means the gluon, whereas the solid line means the quark or antiquark(top).
The differential parton-parton cross section calculated within the LO QCD 
(middle and bottom) \cite{Mangano:2010}} 
% \cite{Mangano:2010}.}
  \end{figure}
Present now the scheme for the calculations of the differential cross section
of the parton-parton elastic scattering within the LO QCD.
Figure 9 (top) illustrates the parton-parton scattering within the LO QCD.
In Fig.9 (middle and bottom) the differential parton-parton cross sections 
calculated within the LO QCD are presented \cite{Mangano:2010}.
%     \begin{figure}[h!]
% %\rotatebox{270}
%  {\epsfig{file=Slaid_14_all.ps,height=12cm,width=14cm  }}
% % {\epsfig{fileAndrei_G.eps=K53p.eps,height=9cm,width=12cm  }}
%  \caption[Fig.9]{The elastic parton-parton scattering within the LO QCD, the wavy line 
%means the gluon, whereas the solid line means the quark or antiquark(top).
%The differential parton-parton cross section calculated within the LO QCD (middle
%and bottom) \cite{Mangano:2010}} 
%% \cite{Mangano:2010}.}
%  \end{figure}
 Here $\Lambda$ is the chromodynamic constant which has taken from 
\cite{PDF:MRST,FF:AKK08}, $n_f$ is the number of flavours, $A_1$ is the dimensional 
coefficient to get the dimension mb/(GeV/c)$^2$ for $d\sigma_{ij}/d{\hat t}$.
The four-momentum transfer squared $Q^2$ is related to the Mandelstam variables
${\hat s},{\hat t},{\hat u}$ for the elastic parton-parton scattering \cite{FFF2} 
\be
Q^2~=~\frac{2{\hat s}{\hat t}{\hat u}}{{\hat s}^2+{\hat t}^2+{\hat u}^2}~.
\label{def:Qsqr}
\ee 
%%%%%%%%%%%%%%%%%%%%%%%%%%%%%%%%%%%%%%%%%%%%%%%%%%%%%%%%%%%%%%%%%%%%%%%%%%%%%%%%%%%%%%%%%%%%%%%%%%%%%%

%%%%%%%%%%%%%%%%%%%%%%%%%%%%%%%%%%%%%%%%%%%%%%%%%%%%%%%%%%%%%%%%%%%%%%%%%%%%%%%%%%%%%%%%%%%%%%%%%%%%%%
 % \begin{figure}[ht]
 %%\rotatebox{270}
 % {\epsfig{file=cylgr.eps,height=3.cm,width=8.cm  }}
 % \caption[Fig.2]{The one-cylinder graph (left pannel) and the multi-cylinder graph (right pannel).} 
 % \end{figure}

\end{document}